# Magnetic Force Microscopy Revealing Molecule Impact on Magnetic Tunnel Junction Based Molecular Devices at Room Temperature

Pawan Tyagi[1,2] and Christopher Riso[1]

*University of the District of Columbia, Department of Mechanical Engineering, 4200 Connecticut Avenue NW Washington DC-20008, USA[1]*
*Email\*: ptyagi@udc.edu*
*University of Kentucky, Chemical and Materials Engineering Department, 177 F Paul Anderson Hall, Lexington, KY-40506, USA[2]*

**Abstract:** Commercially successful magnetic tunnel junction can harness the unmatched capabilities of molecular device elements by solving decade old fabrication issues. Utilization of magnetic tunnel junction as a testbed for molecules also enables unprecedented magnetic studies of molecular spintronics devices. This paper utilizes magnetic force microscopy (MFM) to vividly show that organometallic molecules when bridged between two ferromagnetic electrodes along the magnetic tunnel junction edges, transformed the magnetic electrodes itself. Molecules impacted several hundred-micron areas of ferromagnetic electrodes at room temperature. Complementary, magnetic resonance and magnetometer studies supported the dramatic MFM results. Molecule induced changes in the magnetic electrodes impacted the transport of the magnetic tunnel junction and stabilized as much as six orders smaller current at room temperature. Magnetic tunnel junction based molecular devices can be a gateway to a vast range of commercially viable futuristic logic and memory devices that are controlled by the molecular quantum states near room temperature.

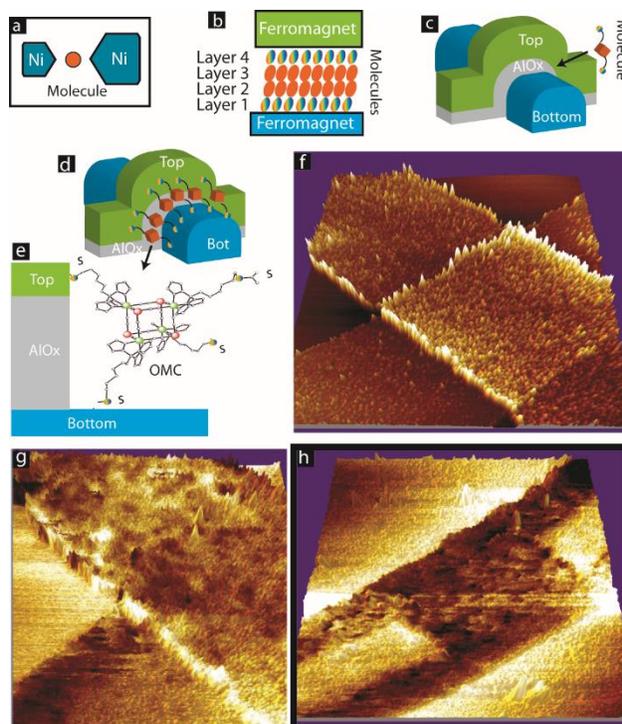

Fig. 1 Molecular spintronics devices based on (a) planar nickel nano-gap junction, (b) multiple molecular monolayers sandwiched between two ferromagnetic electrodes, (c) magnetic tunnel junction with the exposed side edges. (d)Molecules covalently bonded between two ferromagnets. (e)Magnified version of one molecule connected to the ferromagnetic electrodes via thiol groups. (f) Topographical image of a MTJ. MFM of the MTJ (g) before and (h) after the hosting molecular channels along the exposed edges.

**Introduction:** Application of spintronics in the form of magnetic tunnel junction (MTJ), with ferromagnet-insulator-ferromagnet thin film configuration, is currently making a global impact. Transforming this commercially successful MTJs in molecular spintronics devices advances two fields simultaneously[1]. Utilization of molecules can overcome the low spin coherence of the tunnelling barriers and scattering issues at ferromagnet-insulator interfaces[2]. Molecules are unequivocally the most configurable and mass producible nanostructures known to mankind[3]. A mass of review papers have discussed their application in conventional logic and memory devices[2] to quantum computers[4, 5]. To make the best use of molecular quantum states in the spintronics device it is extremely important to directly connect a single molecule to the ferromagnetic electrodes with tunable magnetic properties[1]. It has been shown that MTJ based molecular spintronics devices (MTJMSDs) can provide a robust method of integrating molecular device elements with ferromagnetic electrodes with different magnetic hardness[6]. A mass producible MTJMSD approach addresses the fabrication issues with very low yield planar nano-gap junction approach (Fig. 1a). In a very insightful study Pasupathy et. al.[7] utilized dissimilar nickel (Ni) electrodes to attain ferromagnetic electrodes for a $C_{60}$ molecule based device (Fig. 1a). In another conventional approach, molecular monolayers were self-assembled on ferromagnetic electrodes[8]. Recent spinterface devices showed that molecular layer directly interacting with ferromagnet impact ferromagnetic spin polarization near the surface (Fig. 1b). However, the strong hybridization between the first molecular layer and ferromagnet fades away steeply for other molecular layers (Fig. 1b). Reducing the number of monolayers to enhance the molecular coupling impact or to reduce the defects inherent to multiple molecular layers make the device extremely vulnerable since molecules cannot behave like a sturdy mechanical spacer [9]. An MTJMSD approach address issues with the conventional MSDs shown in Fig1a and b. A prefabricated MTJ (Fig. 1c), where the minimum gap between the top and bottom ferromagnetic electrodes is equal to the tunnelling barrier, can allow the covalent bonding with desired molecules (Fig. 1d). In this approach, one can utilize all the prior MTJ research to tune magnetic electrode properties with angstrom scale precision. Most importantly, the burden of mechanically separating the two ferromagnetic electrodes is borne by the insulating tunnelling barrier, and a molecule covalently bonded to the ferromagnetic electrodes (Fig. 1e) get unprecedented opportunity to display their full potential as a device element. We have previously demonstrated a liftoff based MTJMSD fabrication approach[1, 6]. Here, we discuss the magnetic force microscopy (MFM) and complementary studies to highlight the attributes of integrating molecules along the MTJ edges.

**Experimental details:** These MTJMSDs are formed by covalently bridging the Octametallic Molecular Complex (OMCs) between the top and bottom ferromagnetic electrodes of an MTJ (Fig. 1d-e). An OMC possessed cyanide-bridged Ni and Fe metal ions and $[(pzTp)Fe^{III}(CN)_3]_4$-$[Ni^{II}(L)]_4[O_3SCF_3]_4$ [(pzTp) = tetra(pyrazol-1-yl)borate; L = 1-S(acetyl)tris(pyrazolyl)decane][10] chemical structure. The exposed side edge of an MTJ was produced by the previously published liftoff based molecular device fabrication method[6, 11]. In this study, the MTJ with Ta(5 nm)/Co(5-7 nm)/NiFe(5-3 nm)/AlOx(2 nm)/NiFe (10 nm) configuration was utilized. Utilizing cobalt (Co) in the bottom electrode increased the magnetic hardness of the bottom electrode as compared to top NiFe electrode. An OMC possessed 10 carbon long tethers terminated with the thiol bonds that helped make NiFe –OMC covalent bonding[10, 12]. Details about MTJMSD fabrication[6], OMC attachment[6], and OMC synthesis [12] and characterization has been published elsewhere. A potential MTJ was studied by AFM before (Fig.1f) and after the OMC treatment. For the MFM studies, Molecular Imaging Pico-scan AFM was utilized. A cobalt coated magnetic cantilever was placed 100-200 nm above the surface and was magnetized before MFM studies.

**Results and discussion:** We have explored the impact of covalently bonding an array of the paramagnetic molecules to the two ferromagnetic electrodes of dissimilar magnetic hardness (Fig.1e). The bottom ferromagnet comprises Co and NiFe. Bottom electrode possessed four fold wider hysteresis loop as compared to the top NiFe electrode [11]. The MTJ was studied by AFM before and after transforming it into an MTJMSD. MTJ's topography image was recorded for monitoring structural integrity (Fig. 1f). Corresponding MFM image (Fig.1g) showed the presence of two magnetic electrodes crossing each other at the junction. The MFM image of the top NiFe electrode revealed the microscopic features of the magnetic regions. Interestingly, bridging OMCs between two ferromagnets resulted in the near disappearance of the magnetic signal for the top NiFe electrode in the MTJ vicinity (Fig. 1h). The topography image remained unchanged before and after establishing OMC channels. To further investigate OMC impact MFM was conducted on a number of OMC treated MTJs. On one MTJMSD magnetic signal did not disappear instead the color contrast became very distinct. Top NiFe tend to become light in color and was closer to the background color of the nonmagnetic insulating substrate. However, bottom electrode assumed dark color (Fig. 2a). It is very interesting to see that top and the bottom electrode are becoming radically different, but in the bare state both electrode exhibited similar color (Fig. 1g). The top NiFe electrode also had various yellow-white color pockets evidencing the presence and evolution of new magnetic phases after OMCs connection with the ferromagnetic electrodes. In no other circumstances, such a dramatic contrast in MFM images of an MTJ was observed. It was also noted that an MTJMSD's top magnetic electrode kept changing until it reached a state when the magnetic phase contrast near cross junction was almost unobservable. On another sample fabricated in a different batch a stable disappearance of magnetic contrast on NiFe top electrode was observed (Fig. 2b). However, the neighboring junction on the same sample showed the partial development of the OMC induced phases (Fig. 2c). The MTJMMSD in Fig. 2c showed the disappearance of magnetic phase near the junction and along the edges. In all the four MTJMSD samples discussed in Fig.1 -2, topography images were intact and ensured MTJMSDs were in the physically sound condition. These four MTJMSDs also clearly evidenced that ~10,000 OMCs bridges on an MTJ impacted the large area of the magnetic leads at room temperature. However, it is noteworthy that magnetic leads are several mm long. It is obvious that OMCs impact near MTJMSD junction has to come in equilibrium with respect to the large mass of the magnetic electrode.

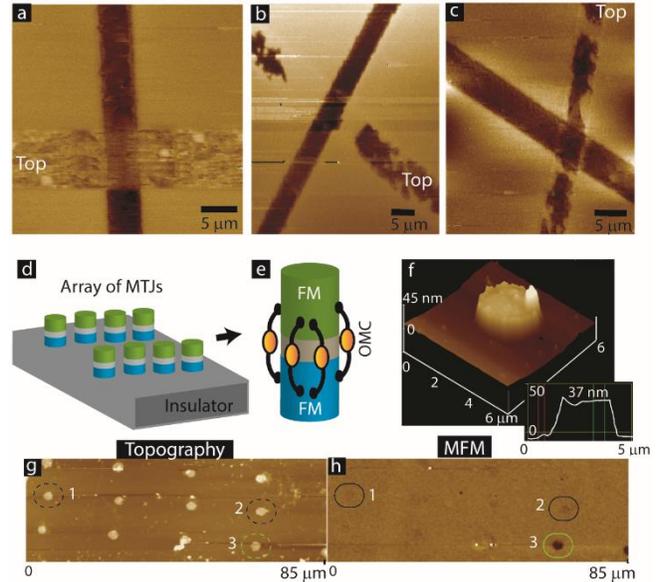

Fig. 2: MFM of MTJMSD (a) with distinct contrast between top and bottom FM electrodes, (b) with vanished magnetic contrast near MTJ cross section, (c) with magnetic contrast in its initial state. (d) An array of cylindrical MTJMSD (e) A cylindrical MTJMSD with OMC molecular channels. (f) Topography of OMC treated MTJ (g) Topographical image of multiple MTJs after interacting with OMCs. (h) MFM images of cylindrical MTJMSD.

To further investigate the OMC impact on the MTJ cross-sectional area an array of 7000 MTJs was produced (Fig. 2d). This array of cylindrical MTJs with the exposed side edges was produced by depositing thin film stack in cylindrical cavities of a photoresist layer and followed by the liftoff. More details about fabrication process mentioned elsewhere [11]. Subsequently, OMC bridges were connected to the top and bottom electrodes to form an array of MTJMSD. This sample preparation strategy is extremely straightforward and avoids the impact of long magnetic leads. Hundreds of MTJMSD were tested for topography study. MTJMSDs were physically intact with an average height of 37 nm (Fig. 2f). On all the MTJMSDs the topography and MFM images were captured simultaneously. Interestingly, the topography of MTJMSD showed the physical presence of MTJMSD but corresponding magnetic contrast disappeared fully on the whole junction at room temperature. For example, the region highlighted by the boxes 1 and 2 showed negligible magnetic contrast. However, occasionally some MTJ appeared with the full contrast. These high magnetic contrast MTJMSD are either defective or failed to host the significant number of OMCs. However, these high contrast MFM images are very useful to confirm that we utilized optimum experimental settings.

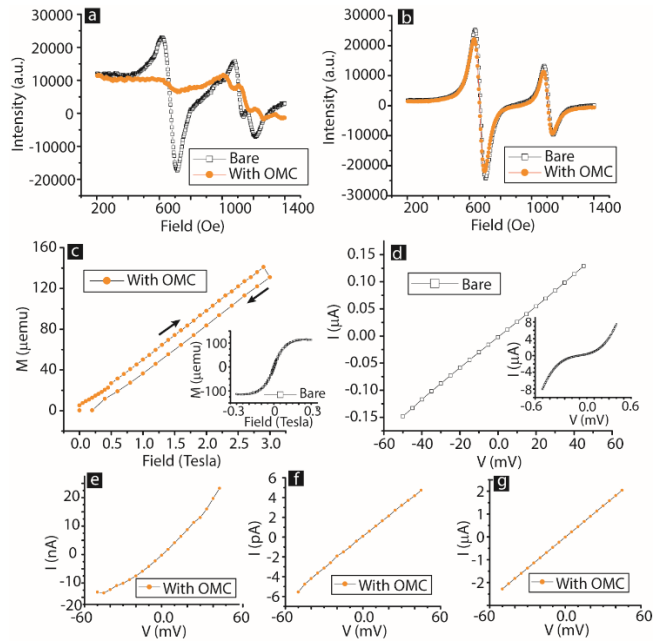

It is logical to think that such a strong OMC impact on MTJs must also be observable with other magnetic measurement methods and transport study. To test this hypothesis, we utilized the array of ~7000 cylindrical MTJs (Fig. 2d) and conducted electron paramagnetic resonance (EPR) and Squid magnetometer study. The first array of MTJ studied by the EPR contained insulating barrier with ~2 nm thickness. Since OMC is ~3 nm long hence it could bridge the insulator gap to transform MTJ into MTJMSD. EPR study clearly showed that bare MTJ possessed two resonance modes corresponding to the system of two ferromagnetic layers separated by a nonmagnetic spacer (Fig.3a). OMCs completely transformed the bare MTJ's resonance modes and asserted with the microscopic impact observed in Fig.1 and Fig. 2. In the control experiment, when an array of MTJ was produced with 4 nm AlOx barrier a negligible change in resonance profile was observed due to OMCs (Fig.3b). This is because of the reason that ~ 3 nm long OMCs were unable to bridge across ~ 4 nm thick AlOx tunnel barrier. This control experiment also provides two additional insights: (a) OMCs interacting or self-assembling on magnetic electrodes is unable to show the impact as it happens when OMC bridge between two ferromagnets. (b) OMCs were harmless to MTJs. In the SQUID magnetometer study, conducted at 150 K to get noise-free signals, an array of ~7000 MTJ showed linear magnetic moment versus magnetic field response (Fig. 3c). Linear magnetic response from MTJMSD was stable when increasing the field to 3T and coming back to zero (Fig. 3c). However, a slight shift in linear response occurred when the field was reducing. Before interaction with OMCs, the array of bare MTJ showed a typical hysteresis loop as shown in the inset graph of Fig. 3c. The linear magnetization curve indicates the presence of either a paramagnetic or two antiferromagnetically coupled ferromagnetic layers. We rule out the possibility of paramagnetic materials because evidently OMCs were coupled to ferromagnetic leads. We found that OMCs developed antiferromagnetic coupling between the top and bottom ferromagnetic electrodes for the cylindrical MTJs. We have discussed this possibility elsewhere [11]. The strength of antiferromagnetic coupling can be estimated by knowing the magnetic field ($H_{ext}$) at which the linear magnetization changed into saturated magnetization state where magnetic moment become invariant of the magnetic field. Since we did not observe the saturation (Fig. 3c), hence we assumed that $H_{ext}$ is at least 3T for the MTJMSD discussed here. Based on the saturation magnetic field antiferromagnetic exchange coupling strength can be expressed by the following expression:

$$J = V.M_a.H_{ext} \tag{1}$$

Fig. 3: Magnetic resonance images of bare MTJMSD (a) with ~ 2 nm AlOx and (b) ~ 4 nm AlOx. (c) Magnetometer study of MTJMSD. Inset shows magnetication loop of the bare MTJ. I-V study of the (d) bare MTJ, inset shows high voltage I-V. I-V of MTJMSD in (e) nA level suppressed state, (f) pA level suppressed state, (g) µA level high current state.

Here, $V$= volume of ferromagnetic electrodes. For the whole array of 7000 MTJs each with 5 µm diameter and 20 nm thick ferromagnetic materials the volume was estimated to be $2.0 \times 10^{-19}$ m$^3$, $M_a$= saturation magnetic field and its magnitude for the NiFe films with which OMCs are directly bonded is $1.2 \times 10^5$ A/m [13]. The magnitude of $J$ with MTJMSD was found to be 22 erg/cm$^2$. On the other hand typical inter-ferromagnetic coupling via the ~2 nm insulator was ~$10^{-3}$ erg/cm$^2$ [14]. This conservative estimation suggested that OMC has increased the exchange coupling between top and bottom ferromagnetic electrodes by as much as five orders of magnitude.

We also studied the impact of OMC induced magnetic ordering and strong exchange coupling on the MTJ's transport. We observed that an MTJ, which showed a typical non-linear tunnelling transport before interacting with OMC (Fig. 3d and the inset graph), generally settled in the suppressed current state. Within first few hours of bridging OMCs between two ferromagnetic electrodes of a cross junction (Fig. 3a) appeared in a suppressed current state below MTJ's leakage current level (Fig. 3e). However, after approximately 24 hours MTJMSD settled in pA current level (Fig. 3f). This phenomenon was observed on many MTJs and suggests that OMC tend to stabilize a suppressed current state. One can perturb the suppressed current state by the application of external voltage and acquire nearly one order higher current above the leakage current level (Fig. 3g). It is noteworthy that OMCs typically increased the tunnel junction current when metallic leads were made up of nonmagnetic materials [6]. OMCs only showed current suppression and magnetic ordering with the magnetic electrodes.

Based on the MTJMSD's I-V studies we estimated the effective barrier heights and barrier thicknesses using Brinkman tunnelling models [15]. A bare MTJ exhibited ~2.2 nm barrier thickness and ~0.7 eV barrier height. After hosting OMC channels along the exposed side edges an MTJ became MTJMSD and showed very different barrier properties in the ~µA range high current state and pA range suppressed current state. According to modeling results in the µA range high current state, an MTJMSD exhibited ~ 1.2 nm barrier thicknesses and ~0.4 eV barrier height. The same MTJMSD in the pA level suppressed current state (Fig. 3) exhibited ~1.4 nm barrier thickness and ~ 2 eV barrier heights. Importantly the barrier thickness after hosting OMCs is equivalent to the length of a decane molecule chain that connected the core of OMC molecules to the NiFe electrode in an MTJMSD (Fig. 4a). The modelled barrier thickness on MTJMSD is also consistent with the barrier thickness obtained on the nonmagnetic tunnel junction.

At this time, we do not have a clear understanding of the underlying mechanism. We provide a tentative mechanism with the help of schematically illustrated Fig. 4. OMC strongly coupled to the MTJ's ferromagnetic electrodes with thiol bonds and low spin scattering alkane tether (Fig. 4a). The MTJMSD provides the best case scenario to realize the impact of OMC channels. OMC's transformative influence is evident from the MFM and other magnetic data (Fig. 1-3) and the estimated increase in inter-ferromagnetic electrode coupling by five orders. In this situation, OMC induced strong coupling and spin transport via alkane tethers to and from OMC cores. Since OMC is paramagnetic hence, it is likely to enable spin filtering and thus strongly affected the spin density of states (Fig. 4b). An OMC is expected to have a net spin state due to the octa-nuclear cage where nickel and iron are bonded via C=N bonds [10]. Due to the selection rule, an OMC in the high spin state is only appearing to host spin down electrons (Fig. 4a). Due to the allowable selective transport of spin down electrons, the spin down density of states of one of the ferromagnetic electrode starts depleting (Fig. 4b). According to MFM study showing it seems top electrode appears to be gaining spin-down electrons because magnetic contrast, the difference between spin- up and spin down density of states, is diminishing (Fig.1). The spin down electrons from bottom electrode start moving into the top electrode near molecule junction vicinity (Fig. 4c). We surmise that the spin-up and spin-down density of states of the top ferromagnetic electrode, around the molecular junction, become equal or comparable. As a result, the net magnetic moment of the top electrode diminishes to zero and appear nonmagnetic (Fig. 4c). However, for the cylindrical MTJMSDs, the redistribution of spin density of states may be different than that from the MTJMSD in cross junction form with extraneous ferromagnetic leads. The area of the OMC impacted regions will be a result of competition between OMC induced exchange coupling induced spin filtering and dimension of the ferromagnetic materials. In the equilibrium state, the new barrier height for spin transport is expected to be very different. For instance, the transport of spin-up electron from the bottom electrode via the dominant OMC channel is expected to be completely blocked. Hence, a high energy barrier was expected (Fig. 4d). However, the barrier height is expected to be smaller between top and electrode and OMC.

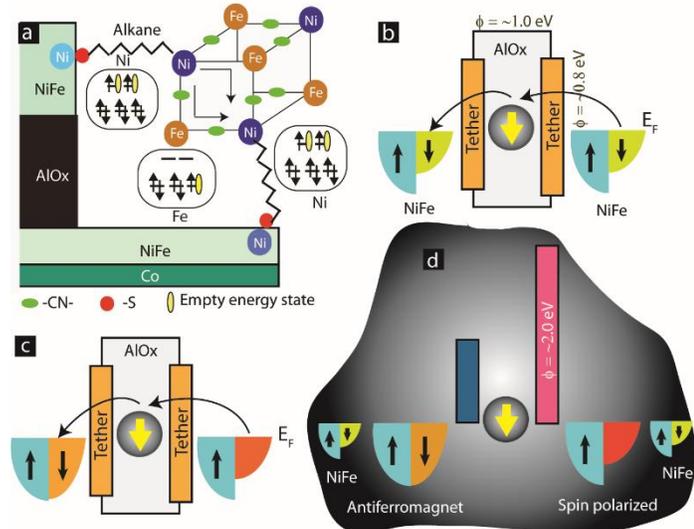

Fig. 4: (a) OMC connected to the NiFe ferromagnets with Ni and Fe spin states. Barrier heights of AlOx and alkane tethers and spin density of states of the MTJMSD (b) before and (c) after spin filtering. (d) Barrier heights between molecular energy states OMC impacted magnetic electrodes.

**Conclusions:**

This paper discussed the experimental magnetic force microscopy (MFM) studies on magnetic tunnel junction based molecular spintronics devices. MFM produced vivid evidence showing that a paramagnetic molecule covalently bonded to the two ferromagnetic electrodes catalyzed a large-scale ordering on ferromagnetic electrodes and impacted several hundred-micron area near molecular junctions. MFM studies were complemented by the magnetometer, EPR, and MFM studies on cylindrical MTJMSDs. Transport studies showed that paramagnetic molecule induced long-range impact on ferromagnetic electrodes resulted in several orders of current suppression at room temperature. Future studies with various forms of magnetic tunnel junctions and molecules are in order. The present work provides the details about the efficacy of cost-effective, and mass producible liftoff based device fabrication. Our study showed that molecules are much more than simple spin channels between two magnetic electrodes. Future study by independent group is necessary to further explore the vast research area of MTJMSD where a large number of MTJs and magnetic molecules can be integrated to attain novel MTJMSDs and magnetic materials.

**Conflicts of interest:** There is no conflict of interest.

**Acknowledgments:** Pawan Tyagi thanks, Dr Bruce Hinds and Department of Chemical and Materials Engineering at the University of Kentucky for facilitating experimental work on MTJMSD during his PhD. OMC was produced Dr Stephen Holmes's group. The preparation of this paper and supporting studies were in part supported by National Science Foundation-Award (Contract # HRD-1238802), Department of Energy/National Nuclear Security Agency (Subaward No. 0007701-1000043016), and Air Force Office of Sponsored Research (Award #FA9550-13-1-0152). Any opinions, findings, and conclusions expressed in this paper are those of the author(s) and do not necessarily reflect the views of any funding agency and authors' affiliations.